\documentclass[aps,prl,preprint,superscriptaddress,notitlepage]{revtex4-1}
\usepackage{times}
\usepackage{amsmath}
\usepackage{bbold}
\usepackage{graphicx}
\usepackage{epstopdf}
\usepackage{upgreek}
\usepackage[english]{babel}
\usepackage{color}
\usepackage{bm}
\usepackage[colorlinks]{hyperref}
\hypersetup{
colorlinks=true,
linkcolor=blue,          % color of internal links
citecolor=blue,          % color of links to bibliography
filecolor=magenta,       % color of file links
urlcolor=cyan
}

\usepackage{dcolumn}% Align table columns on decimal point
\usepackage[normalem]{ulem}
\usepackage[colorinlistoftodos,prependcaption,textsize=tiny]{todonotes}

\DeclareFontFamily{U}{mathx}{\hyphenchar\font45}
\DeclareFontShape{U}{mathx}{m}{n}{<-> mathx10}{}
\DeclareSymbolFont{mathx}{U}{mathx}{m}{n}
\DeclareMathAccent{\widebar}{0}{mathx}{"73}

\newcommand{\yig}{\textsc{yig}}

\begin{document}

%\preprint{AIP/123-QED}

\title{Low-impedance superconducting microwave resonators for strong coupling\\ to small magnetic mode volumes}

\author{L. McKenzie-Sell}
	\email{lam98@cam.ac.uk}
\affiliation{Department of Materials Science and Metallurgy, University of Cambridge, Cambridge, CB3 0FS, United Kingdom}
\affiliation{Cavendish Laboratory, University of Cambridge, Cambridge, CB3 0HE, United Kingdom}

\author{J. Xie}
\affiliation{Cavendish Laboratory, University of Cambridge, Cambridge, CB3 0HE, United Kingdom}

\author{C.-M. Lee}
\affiliation{Department of Materials Science and Metallurgy, University of Cambridge, Cambridge, CB3 0FS, United Kingdom}

\author{J. W. A. Robinson}
\affiliation{Department of Materials Science and Metallurgy, University of Cambridge, Cambridge, CB3 0FS, United Kingdom}

\author{C. Ciccarelli}
\affiliation{Cavendish Laboratory, University of Cambridge, Cambridge, CB3 0HE, United Kingdom}

\author{J. A. Haigh}
	\email{jh877@cam.ac.uk}
\affiliation{Hitachi Cambridge Laboratory, Cambridge, CB3 0HE, United Kingdom}

\date{\today}

\begin{abstract}
Recent experiments on strongly coupled microwave and ferromagnetic resonance modes have focused on large volume bulk crystals such as yttrium iron garnet, typically of millimeter-scale dimensions. We extend these experiments to lower volumes of magnetic material by exploiting low-impedance lumped-element microwave resonators. The low impedance equates to a smaller magnetic mode volume, which allows us to couple to a smaller number of spins in the ferromagnet. Compared to previous experiments, we reduce the number of participating spins by two orders of magnitude, while maintaining the strength of the coupling rate. Strongly coupled devices with small volumes of magnetic material may allow the use of spin orbit torques, which require high current densities incompatible with existing structures.

\end{abstract}

\maketitle

The coherent transformation of the excitations of a condensed matter system to an electromagnetic mode is made possible when they are strongly coupled \cite{kimble_strong_1998}, i.e. the dissipation rates of the two systems are less than their coupling rate. This is typically harder to achieve through magnetic than electric fields due to the lower interaction strength in free space \cite{jackson_classical_1998}. However, the relative weakness of the magnetic dipole moment can be compensated for by the $\sqrt{N}$ enhancement \cite{soykal_strong_2010} of the coupling rate with the number of spins $N$. This has allowed the strong coupling of microwave and ferromagnetic resonance modes, recently explored for quantum magnonic \cite{lachance-quirion_resolving_2017} and microwave-optical conversion \cite{hisatomi_bidirectional_2016} applications. While the microwave resonance has been coupled to millimeter-scale bulk yttrium iron garnet (YIG) crystals \cite{Huebl_High_2013,tabuchi_hybridizing_2014,zhang_strongly_2014,goryachev_high-cooperativity_2014}, or large scale thin films \cite{bai_spin_2015}, reliance on increased $N$ to enhance coupling cannot extend to micro- and nanoscale devices.

The strength of dipole coupling rates in device structures can also be significantly modified through the impedance of the electromagnetic mode \cite{devoret_circuit-qed:_2007}. Lowering or raising the impedance tunes the relative amplitudes of the magnetic and electric field fluctuations, which can be chosen as appropriate for the specific application. This idea has enabled the optimization of coupling rates of microwave resonators to superconducting qubits \cite{bosman_approaching_2017} and semiconductor quantum dots \cite{stockklauser_strong_2017}, and is driving the development of novel low-impedance resonators \cite{eichler_electron_2017}, which should allow inductive electron spin resonance experiments on a single atom \cite{haikka_proposal_2017}. 

\begin{figure}%
\includegraphics[width=0.5\columnwidth]{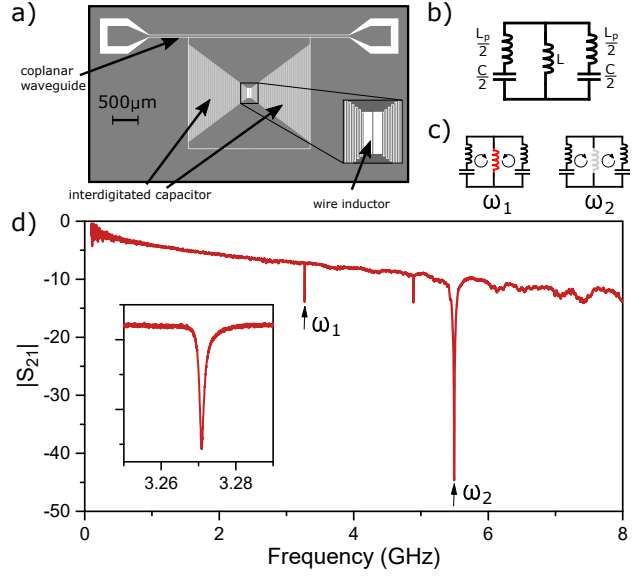}%
\caption{(a) Schematic of the resonator design. (b) Equivalent circuit which reproduces the behavior of the first two modes. (c) Diagram indicating current flow in the fundamental (left) and second (right) harmonic modes. For the second harmonic there is no current flow through the wire inductor. (d) Transmission spectrum of the resonator without YIG layer. Inset shows detail of the fundamental mode.}
\label{fig:fig1}%
\end{figure}

In this letter, we demonstrate that low-impedance resonators \cite{eichler_electron_2017} can compensate for the reduction in $N$, to reach strong coupling of magnons and microwave photons at much smaller magnetic mode volumes than has been achieved previously. This may lead to novel spintronics effects, such as spin-orbit torques, playing a role in strongly-coupled devices. In particular, anti-damping torques \cite{brataas_current-induced_2012} can be used to modify the linewidth of ferromagnetic resonance modes \cite{demidov_control_2011}, which could be useful for exploring exceptional point physics \cite{zhang_observation_2017,grigoryan_synchronized_2018}, or driving auto-oscillations \cite{demidov_magnetic_2012,duan_nanowire_2014} in this hybrid system.

We take the design of these low-impedance resonators from recent progress towards electron spin resonance on a single spin of an isolated donor in silicon \cite{bienfait_reaching_2016,eichler_electron_2017}. These lumped element resonators consist of a small inductor shunted by a large capacitance (leading to low impedance $Z=\sqrt{L/C}$). The small inductor localizes the magnetic field of the mode. The volume of magnetic material interacting with the microwave mode is therefore reduced without micro-patterning of the YIG film. This avoids the issues that can arise from patterning yttrium iron garnet, for example by etching \cite{heyroth_monocrystalline_2018}, which in general introduces roughness or material damage that increases the damping rates of ferromagnetic resonance modes.

A schematic of the microwave resonator is shown in Fig.\,1, following the design in Ref.\,\onlinecite{eichler_electron_2017}. The inductor $L$ is a small strip-line of length $l_\text{ind}$, and the capacitor $C$ is an interdigitated finger structure distributed on both sides of the inductor. The resonator is driven and measured through a coplanar waveguide to which the resonator is coupled. The resonators are fabricated from 80-nm-thick Nb, which is magnetron sputtered onto single crystal sapphire, by optical lithography and reactive ion etching. Electrical measurements are performed %
%at 4.2~K %
in a helium bath cryostat, with a single axis superconducting magnet for applying magnetic fields. The transmission $|S_{21}|$ of the coplanar waveguide is measured with a vector network analyzer.

We first characterize the superconducting microwave resonators. The fundamental frequency of the resonator is measured to be $3.27$\,GHz, with the second harmonic at ${\omega_2}/({2\pi}) = 5.49$\,GHz, in reasonable agreement with calculated values \cite{supplementary_info}. The loaded Q-factor of the fundamental mode is $Q_L\approx1400$, while that of the second harmonic is $Q_L\approx40$, due to significant loading from the large inductive coupling to the line.

\begin{figure}%
\includegraphics[width=0.5\columnwidth]{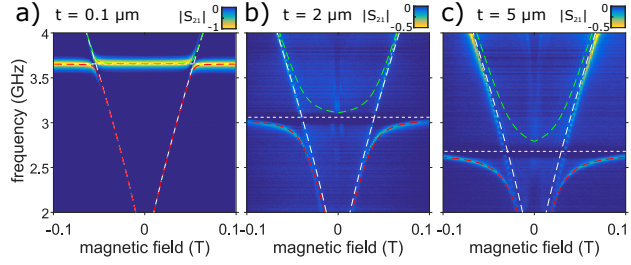}%
\caption{Transmission of the coplanar waveguide as a function of applied magnetic field and frequency, for YIG layers of (a) 0.1\,$\mu$m (b) 2\,$\mu$m, and (c) 5\,$\mu$m. A field independent background has been subtracted. The white lines indicate the bare microwave resonator (short dash) and ferromagnetic resonance (long dash) frequencies. The green and red dashed lines are the coupled modes fitted to the measured data.}%
\label{fig:fig2}%
\end{figure}

The parasitic inductance is estimated from the frequency of the second harmonic $\omega_2 = 1/\sqrt{L_p C/4}$ [see Fig.\,1(b)], using $C = 4.7$\,pF, from the analytical expression for the capacitance of an interdigitated capacitor \cite{igreja_analytical_2004} (adapted to the trapezoidal geometry \cite{supplementary_info}). This gives $L_p \approx 0.715$\,nH. This then allows us to calculate the inductance of the inductive wire, using the fundamental frequency $\omega_1 = 1/\sqrt{(L_0 + L_p/4)C}$, giving $L_0\approx 0.325$\,nH. We therefore estimate the impedance of the resonator to be $Z_R = \sqrt{(L_0 + L_p/4)/C} \approx 10.4$\,$\Omega$, significantly lower than the typical characteristic impedance $Z = 50$\,$\Omega$ and that of free space $Z = 377$\,$\Omega$. We find similar results for finite element device modeling \cite{supplementary_info}. The resonator is found to be robust to in-plane magnetic fields up to $\sim0.5$\,T.

We now characterize the coupling of the microwave resonator to YIG thin films. In these measurements we concentrate on the fundamental mode, in which current density is maximized in the inductor wire, localizing the magnetic field. We use three YIG layers of thickness $d_\yig=0.1,2,5$\,$\mu$m, grown by liquid phase-epitaxy on gadolinium gallium garnet substrates \cite{dubs_sub-micrometer_2017}. These layers are successively mounted face-down on the microwave resonator, and held in place by vacuum grease. 

\begin{figure}%
\includegraphics[width=0.4\columnwidth]{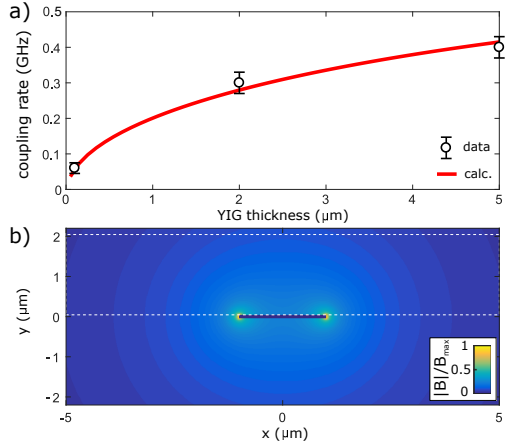}%
\caption{(a) Measured coupling rate verses YIG thickness, with calculated fit (red line). (b) Magnetic field distribution around the inductor strip, used to fit the coupling rate with Eq.~\ref{eq:eta2approx}. The dashed white lines indicate where the upper and lower surfaces of a 2\,$\mu$m thick YIG film would lie.}%
\label{fig:fig3}%
\end{figure}

Measurements of the coplanar waveguide transmission $|S_{21}|$ as a function of magnetic field are shown in Fig.\,2, for the three film thicknesses, where the YIG layer covers the entire area of the resonator. The magnetic field is applied in the plane of the resonator, along the direction of the inductive wire, such that the microwave magnetic field is perpendicular to it. Firstly, it can be seen that loading the resonator with the YIG film affects the resonance frequency, as both $\mu$ and $\epsilon$ are modified: this results in a frequency shift of the resonator with increasing YIG thickness. Secondly, for all thicknesses there is clear normal mode splitting of the resonator modes with the ferromagnetic resonance mode. However, for the two thicker layers, only one branch of the coupled modes is observed. We believe that this could occur for two reasons. Firstly, the symmetry of the drive field can affect which modes are driven \cite{lambert_cavity-mediated_2016}. Secondly, the damping of the upper branch may be much larger as it overlaps with the spin-wave band of the YIG \cite{roberts_magnetodynamic_1962}, meaning that it is difficult to observe.

The coupling rate for the three layers is extracted using a manual fit to the lower branch. The magnetic mode frequency is described by the Kittel formula $\omega_m = \gamma\mu_0 \sqrt{H_0 (H_0 - M)}$, where $H_0$ is the applied magnetic field, $M$ is the effective magnetization of the YIG, and $\gamma/(2\pi)=28$\,GHzT$^{-1}$ is the gyromagnetic ratio. We use a simple coupled mode model of $\omega_m$ and the microwave resonator mode $\omega_c$, so that the hybrid modes are given by $\omega_{\pm}= (\omega_m + \omega_c)/2 \pm \sqrt{g^2 + ((\omega_m - \omega_c)/2)^2}$. The extracted coupling rate $g$ is plotted in Fig.\,3, where a sub-linear increase with thickness is observed.

We can understand this trend by considering the theoretically expected coupling rate,
\begin{equation}
g = \frac{\eta}{2} \gamma  \sqrt{\hbar \mu_0 \omega} \sqrt{{\rho s}},
\label{eq:g}
\end{equation}
where $\rho$ is the density of spins $s$. The rate is proportional to the mode overlap of the magnon $\vec{m}$ and microwave $\vec{B}$ magnetic fields \cite{lambert_identification_2015},
\begin{equation}
\eta^2 = \frac{\left|\int{\delta\vec{m}^{}(\vec{r})\cdot\vec{B}^*(\vec{r})}dV\right|^2}{\int{|\delta\vec{m}^{}(\vec{r})|^2}dV\int{|\vec{B}(\vec{r})|^2}dV},
\label{eq:eta2}
\end{equation}
which will depend on how much of the microwave magnetic field mode volume is filled with magnetic material \cite{Huebl_High_2013}. Note that although the impedance does not appear in this expression, it affects the rate implicitly through this mode volume.

The magnon mode function is assumed to be Kittel-like, i.e. with uniform amplitude and phase of the ferromagnetic resonance oscillation, but localized around the wire. We approximate its spatial extent by a cuboid function, with two of the dimensions given by the thickness of the film and the length of the inductor wire. The third dimension is the distance across the film, perpendicular to the inductor, over which Kittel-like ferromagnetic resonance is excited by the inductor. This extent is left as a fitting parameter which will be determined from the coupling rate. This allows us to approximate $\eta$ by evaluating
\begin{equation}
\eta^2 \approx \frac{\left|\int_{V_M}{{B_x}(\vec{r})}dV\right|^2}{V_M \int{|\vec{B}(\vec{r})|^2}dV},
\label{eq:eta2approx}
\end{equation}
where the magnon mode volume is $V_M\approx l_\text{ind}d_\yig w_{\delta\vec{m}}$. We evaluate the numerator by considering the magnetic field distribution in the 2D plane perpendicular to the wire using the approximate distribution of current in a superconducting strip \cite{duzer_principles_1998}. This field distribution is shown in Fig.\,3(b). The integral in the denominator is evaluated from the same field distribution, but must be corrected by a factor $L_0/(L_0+ L_p/4)$ to account for the magnetic field due to parasitic inductance in the device. The effect of kinetic inductance, negligible for our film \cite{supplementary_info}, was omitted. However, we have also performed measurements with high kinetic inductance niobium nitride thin films \cite{annunziata_tunable_2010}, in which the coupling rates are reduced by the significant additional inductance of the strip-line.

The only free parameter in the calculation is the extent of the magnon mode in the transverse direction $w_{\delta\vec{m}}$. By adjusting this parameter, we obtain reasonable agreement with the experimental data, as shown by the solid line in Fig.\,3(a). The value is $w_{\delta\vec{m}}/2\approx15$\,$\mu$m. This length-scale can be understood as the spin-wave propagation length, or the coherence length of the uniform mode, and is in reasonable agreement with other experiments in YIG thin films \cite{chumak_magnon_2014}.

The analysis above assumes that the coupling was localized around the inductive wire, as that is where the microwave magnetic field is concentrated. However, although the current density is weaker elsewhere in the resonator \cite{supplementary_info}, the interaction volume is much larger, so in principle could contribute significantly to the coupling rate. To characterize this contribution, we performed further measurements using a small 2\,$\mu$m thick YIG chip (see Fig\,.4(b)). With this smaller piece we can cover the inductor wire, while minimizing the coverage of the interdigitated capacitor, as shown schematically in Fig.\,4(b). In this measurement we see that the coupling rate is $g/(2\pi)=380$\,MHz, compared to $g/(2\pi)=300$\,MHz when the 2\,$\mu$m YIG layer covers the entire resonator. That these values are comparable indicates that most of the coupling indeed comes from a localized area around the inductive wire.

From our experiments, the total volume of magnetic material contributing to the normal mode splitting is estimated to be $V_M\approx1.2\times10^{-14}$~m$^3$. Given the spin density $\rho=2.1\times10^{22}$~cm$^{-3}$, the corresponding number of spins $N\approx2.6\times10^{14}$ is two orders of magnitude smaller than that achieved in YIG with a standard coplanar waveguide resonator \cite{Huebl_High_2013}. 
\begin{figure}%
\includegraphics[width=0.5\columnwidth]{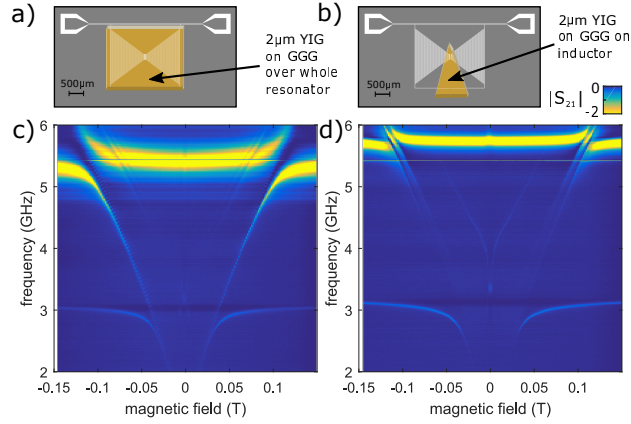}%
\caption{Comparison of measurements with (a) YIG layer covering the entire chip and (b) YIG layer largely covering only the inductor. A field independent background has been subtracted. Note that avoided crossings of the fundamental and second-harmonic resonator modes are shown.}%
\label{fig:fig4}%
\end{figure}

We have shown that low-impedance superconducting microwave resonators promise an excellent path towards strong coupling of microwave fields to magnetization dynamics at the nano-scale. This circumvents the need for high resolution lithography of magnetic elements, which typically leads to degradation of the ferromagnetic resonance properties. The use of this type of resonator may be transferable to similar experiments in magnetic molecules \cite{mergenthaler_strong_2017}, and in achieving strong-coupling to metallic ferromagnet thin films. Although the linewidth of the ferromagnetic resonance in metallic layers is not as narrow as in YIG, the application of anti-damping spin-orbit torques \cite{brataas_current-induced_2012} may allow this problem to be overcome. 

As well as applications that require strong coupling at the micron-scale, in the weak coupling limit these types of resonators may allow direct inductive ferromagnetic resonance experiments on small magnetic objects. Typically, ferromagnetic resonance must be measured on an array of such devices to attain sufficient signal, but device inhomogeneity can then have a significant impact on the measured linewidth \cite{heyroth_monocrystalline_2018}. Alternatively, non-inductive readout mechanisms can be used, such as optical \cite{hiebert_direct_1997}, torsional \cite{losby_torque-mixing_2015}, or resistive \cite{costache_large_2006} methods, but these add additional measurement complications. 

Finally, recent efforts towards microwave-optical conversion in yttrium iron garnet \cite{hisatomi_bidirectional_2016,zhang_optomagnonic_2016,osada_cavity_2016,haigh_selection_2018} have been limited by the poor overlap of magnetic and optical modes. By coupling microwaves into a much smaller volume of magnetic material, it should be possible to increase the conversion-efficiency significantly \cite{viola_kusminskiy_coupled_2016}.  While optical cavities of wavelength $\lambda\sim1$~$\mu$m can be made with mode volume $V \sim 1$\,$\mu$m$^3$\,\cite{dolan_femtoliter_2010}, coupling microwaves to that volume may require novel techniques such as those outlined here.

\section*{Acknowledgments}
\begin{acknowledgments}
We are grateful for useful discussions with Andreas Nunnenkamp, Imtiaz Ahmed, Fernando Gonzalez-Zalba and Silvia Viola-Kusminsky, as well as NbN films grown by Yi Zhu. This work was supported by the European Union’s Horizon 2020 research and innovation programme under grant agreement No 732894 (FET Proactive HOT). C.C. acknowledges the Royal Society and the Winton Trust. J.W.A.R. and L.M-S. acknowledge funding from the Royal Society and the EPSRC through the \emph{International network to explore novel superconductivity at advanced oxide superconductor/magnet interfaces and in nanodevices} (EP/P026311/1) and Programme Grant \emph{Superspin} (EP/N017242/1). L.M-S. also acknowledges the Winton Trust. 
\end{acknowledgments}
%\bibliography{bibliography}
%apsrev4-2.bst 2019-01-14 (MD) hand-edited version of apsrev4-1.bst
%Control: key (0)
%Control: author (8) initials jnrlst
%Control: editor formatted (1) identically to author
%Control: production of article title (0) allowed
%Control: page (0) single
%Control: year (1) truncated
%Control: production of eprint (0) enabled
%

%
\end{document}